\documentclass[a4paper,12pt]{article}
\pdfoutput=1
\usepackage{amssymb}
\usepackage{amsmath}
\usepackage{amsfonts}
\usepackage{mathtools}
\usepackage[noend]{algpseudocode,algorithm}
\usepackage{slashed}
\usepackage{color}
\usepackage{indentfirst}
\usepackage{graphicx}
\usepackage{bbm}
\usepackage{csquotes} 
\usepackage{caption}
\usepackage{cite}
\usepackage{here}
\usepackage{comment}
\usepackage{listings}
\makeatletter
\def\BState{\State\hskip-\ALG@thistlm}
\makeatother
\makeatletter
\renewcommand{\ALG@name}{Algorithm}
\makeatother
\algrenewcommand\algorithmicdo{}

\newcommand{\tr}{\mathrm{tr}}
\newcommand{\figref}[1]{Fig.\ref{fig:#1}}
\renewcommand{\algref}[1]{Algorithm \ref{#1}}

\renewcommand{\thefootnote}{\fnsymbol{footnote}}

\begin{document}

\title{
\begin{flushright}
\begin{minipage}{0.2\linewidth}
\normalsize
KEK-TH-2200 \\
WU-HEP-20-4\\*[50pt]
\end{minipage}
\end{flushright}
{\Large \bf 
Deep learning and k-means clustering\\in heterotic string vacua with line bundles\\*[20pt]}}

\author{Hajime~Otsuka$^{a}$\footnote{
E-mail address: hotsuka@post.kek.jp
}
\ and \
Kenta~Takemoto$^{b}$\footnote{
E-mail address: phys-tm.3000@asagi.waseda.jp
}\\*[20pt]
$^a${\it \normalsize 
KEK Theory Center, Institute of Particle and Nuclear Studies, KEK,}\\
{\it \normalsize 1-1 Oho, Tsukuba, Ibaraki 305-0801, Japan}\\
$^b${\it \normalsize 
Department of Physics, Waseda University, Tokyo 169-8555, Japan}} 
\maketitle

\date{
\centerline{\small \bf Abstract}
\begin{minipage}{0.9\linewidth}
\medskip 
\medskip 
\small
We apply deep-learning techniques to the string landscape, 
in particular, $SO(32)$ heterotic string theory on 
simply-connected Calabi-Yau threefolds with line bundles. 
It turns out that three-generation models cluster in particular islands specified by deep autoencoder networks and k-means++ clustering. 
Especially, we explore mutual relations between model parameters and the cluster with densest three-generation models (called ``3-generation island").
We find that the 3-generation island has a strong correlation with the topological data of Calabi-Yau threefolds, 
in particular, second Chern class of the tangent bundle of the Calabi-Yau threefolds. 
Our results also predict a large number of Higgs pairs in the 3-generation island.
\end{minipage}
}

\renewcommand{\thefootnote}{\arabic{footnote}}
\setcounter{footnote}{0}
\thispagestyle{empty}
\clearpage
\addtocounter{page}{-1}

\tableofcontents

\section{Introduction}\label{sec:introduction}

The deep learning is an attractive method not only for 
image recognition, but also for applications to string theory as well 
as particle physics. 
So far, neural networks have been applied to explore the 
vacuum structure of string theory~\cite{He:2017aed,He:2017set,Krefl:2017yox,Ruehle:2017mzq}, 
for instance, 
a conjecture for the gauge group rank in F-theory compactifications~\cite{Carifio:2017bov}, 
identification of fertile islands in the toroidal orbifold landscape \cite{Mutter:2018sra}, 
the prediction of the Hodge numbers of Calabi-Yau (CY)  manifolds~\cite{Bull:2019cij}, exploring Type IIA compactifications with intersecting D6-branes \cite{Halverson:2019tkf}, landscape of Type IIB flux vacua~\cite{Cole:2019enn} and $E_8\times E_8$ heterotic line bundle models~\cite{Larfors:2020ugo}, and finding the numerical 
CY metric \cite{Ashmore:2019wzb}. 
(For more details, see, Ref.~\cite{Ruehle:2020jrk}.) 
Utilizing machine learning techniques for exploring 
the Standard Model (SM) vacua from string theory will give new 
insights into the string landscape. 

Among the huge number of CY manifolds, there 
exists a restricted class of non-simply connected 
CYs where Wilson lines can be introduced to obtain 
the SM gauge group. 
To search for the Minimal Supersymmetric SM (MSSM) on a vast 
class of CYs, a hypercharge flux breaking scenario 
is proposed in Refs.~\cite{Beasley:2008kw,Donagi:2008kj,Blumenhagen:2005ga,Blumenhagen:2005pm,Blumenhagen:2006ux,Blumenhagen:2006wj,Otsuka:2018rki}. 
Such a direct flux breaking scenario does not require an 
existence of Wilson lines and is applicable to a vast class 
of CYs. 
However, in the usual random scan in the string landscape, it is difficult to reveal the  origin of the MSSM-like models from a huge number of compactification parameters. To resolve this issue, we apply the machine learning technique for an exhaustive search of 
four-dimensional (4D) string models based on the direct flux breaking scenario.  
It opens up a new possibility of searching for the three-generation SM. 
Especially, we follow the systematic approach proposed in Ref.~\cite{Otsuka:2018rki} 
on the basis of $SO(32)$ heterotic string theory with line bundles. 
The direct flux breaking scenario in $SO(32)$ heterotic string theory is motivated by 
its S- and T-dual intersecting D6-brane models in Type IIA string theory, 
where several stacks of D-branes directly lead to the MSSM-like 
gauge group. 

The purpose of this paper is to reveal the nature of SM vacua in $SO(32)$ heterotic line bundle models by employing the neural network technique, in particular the origin of three generations of quarks and leptons. 
So far, the machine learning techniques have been utilized to reproduce known physical quantities, like topological 
data and numerical metric of CYs, but our approach attempts to not only reproduce the known physical data, but also find a characteristic feature of the CY compactifications. 

In this paper, we adopt a deep autoencoder~\cite{Hinton} to find the parameter region leading to the 
SM-like spectra on CY threefolds, in particular, Complete Intersection CY (CICY)~\cite{Candelas:1987kf,Candelas:1987du}.\footnote{For CICY lists, see, \cite{CICY}.} The autoencoder is useful to reduce the higher-dimensional parameter space to the 2D one by extracting characteristic features of the data, although the neutral network itself does not have the knowledge of SM spectra. To classify the characteristic features from the autoencoder, we deal with the k-means++ clustering~\cite{KM++}. 
After performing the autoencoder and the k-means++ clustering to the vacua, we find that three-generation models are clustered in particular islands, in a similar to 
the analysis in the toroidal orbifold models \cite{Mutter:2018sra} and we call the cluster 
with densest three-generation models ``3-generation island". 
By introducing Kullback–Leibler (KL) divergence \cite{KL}, we also find that the 3-generation island is strongly correlated with the topological data of CY, in particular, 
the second Chern number of CY threefolds.
Our approach enables us to capture the nature of not only the known toroidal orbifold landscape but also a large class of CY compactifications. 

This paper is organized as follows. 
In Section \ref{sec:2}, we briefly review the phenomenological and theoretical 
constraints in $SO(32)$ heterotic line bundle models to implement the deep autoencoder. 
In Section \ref{sec:3}, we show the algorithms of the autoencoder and k-means++ clustering utilized in 
the dataset of heterotic line bundle models.  As discussed in Section \ref{sec:4}, we find that $n$-generation models are clustered in the 2D space derived by the autoencoder. 
Especially, we focus on the 3-generation island 
and extract its phenomenological consequence. 
Section 5 is devoted to the conclusions and discussions. 

\section{Line bundle models}\label{sec:2}

Before applying the deep autoencoder to the 
dataset of heterotic line bundle models, we briefly review the heterotic 
line bundle models developed in Refs.~\cite{Blumenhagen:2005ga,Blumenhagen:2005pm,Otsuka:2018rki,Anderson:2011ns,Anderson:2012yf,Abe:2015mua,Otsuka:2018oyf} with an 
emphasis on the phenomenological and theoretical constraints. 

\subsection{Consistency conditions}
\label{sec:2_1}

We start from the low-energy effective action of $SO(32)$ heterotic string theory on smooth CY manifolds with 
multiple line bundles.\footnote{The following discussion 
is applicable to the $E_8\times E_8$ heterotic string theory as well.}
The total internal gauge bundle consists of the 
multiple internal line bundles $L_a$ with structure group $U(1)$, namely
\begin{align}
    W =\bigoplus_{a} L_a,
\end{align}
where $U(1)$ is supposed to descend from $U(N) \subset SO(2N) \subset SO(32)$ and the concrete embedding into $SO(32)$ is shown later. 
In contrast to the standard embedding scenario, such a 
non-standard embedding has a phenomenological and theoretical rich structure. 
For instance, $SO(32)$ gauge group is broken down to the 4D gauge group $G$ rather than $E_6$ because of the non-vanishing background field values of the $SO(32)$ gauge field strength. 
The introduction of the internal line bundles each with structure group $U(1)$ corresponds to an existence of  internal $U(1)_a$ gauge fluxes $F_a$,
\begin{align}
    {\rm tr} (F_a) = 2\pi \sum_{i=1}^{h^{1,1}} {\rm tr}(T_a) m_a^i w_i,
\end{align}
where the $m_a^i$ are flux quanta in the basis of K\"ahler form $w_i$, $i=1,2,\cdots, h^{1,1}$ with $h^{1,1}$ being the Hodge number of CY ${\cal M}$. Here $T_a$ are $U(1)_a$ generators and ``tr" represents for the trace in the fundamental representation. 
We remark that the internal gauge fluxes are now turned on only $H^{1,1}({\cal M})$, otherwise the 
supersymmetry is broken by the $F$-term in the 4D effective action. Even when the fluxes 
are restricted on $H^{1,1}({\cal M})$, it is required to check the $D$-term condition (zero-slope poly-stability condition) for each $U(1)_a$. Since it depends on the structure of K\"ahler cone and the $D$-term conditions give rise to Diophantine equations, we do not take into account the $D$-term condition in the neural network. It is possible to check the $D$-term conditions for each model after obtaining the three-generation models.\footnote{For the discussion of Diophantine equations in 
Type IIA intersecting D6-brane models, see, Ref.~\cite{Halverson:2019tkf}. We leave the implementation of such Diophantine equations to the neural network for future work.} 
As a consequence of the above gauge fluxes, we obtain 
not only the SM-like gauge group, but also the chiral fermions. 
Before going to discuss the phenomenological models, we enumerate the theoretical constraints for gauge fluxes. 
We will take into account these constraints when we implement the line bundle models to the deep neural networks. 
\begin{enumerate}
    \item Tadpole cancellation condition
    
    Internal gauge fluxes cause the gauge and gravitational anomalies in the 4D effective action. To cancel these anomalies, we require the cancellation of the tadpole for the Neveu-Schwarz (NS) sector,
    \begin{align}
        {\rm ch}_2 (W) +c_2(T{\cal M}) =\left( \frac{1}{2}\sum_{a,i,j}{\rm tr}T_a^2m_a^i m_a^j d_{ijk} +c_{2,k}\right)\hat{w}^k =N_k \hat{w}^k,
    \end{align}
    where we expand the second Chern character of $W$ and second Chern class of the tangent bundle of CY ${\cal M}$ in the basis of $\hat{w}^k \in H^{2,2}({\cal M})$ satisfying
    \begin{align}
        \int_{{\cal M}} \hat{w}^k\wedge w_i =\delta^k_i.
    \end{align}
    We denote the triple intersection number of CY by $d_{ijk}=\int_{{\cal M}}w_i\wedge w_j \wedge w_k$ and we introduce the NS5-branes wrapping holomorphic two-cycles. To ensure the stability of our system, we prohibit an existence of anti NS5-branes, namely
    \begin{align}\label{eq:NS5 condition}
        N_k =\frac{1}{2}\sum_{a,i,j}{\rm tr}T_a^2m_a^i m_a^j d_{ijk} +c_{2,k} \geq 0.
    \end{align}
    \item K-theory condition
    
    Next, we require the existence of spinor bundles on CY manifolds, corresponding to the requirement of the trivial Stiefel-Whitney class,
    \begin{align}\label{K-theory condition}
        c_1 (W) \in H^{1,1}({\cal M}, 2\mathbb{Z}) \leftrightarrow {\rm tr}T_a m_a^i \equiv 0\,\, ({\rm mod}\,2) 
    \end{align}
    for $i=1,2,\cdots, h^{1,1}$. This condition is related to K-theory condition developed in the S-dual Type I string theory~\cite{Witten:1998cd,Uranga:2000xp}.
    In some cases, we replace \eqref{K-theory condition} with
    \begin{equation}\label{simplified K-theory condition}
        \tr T_a m_a^i = 0
    \end{equation}
    for simplicity because of the limitation of the computing power.
    \item $U(1)_Y$ masslessness conditions
    
    Non-trivial gauge background induces the 4D St\"uckelberg couplings between string axions associated with Kalb-Ramond $B$-field and $U(1)$ gauge bosons  through Green-Schwarz terms. Indeed, it causes the mass terms for some anomalous $U(1)$s~\cite{Blumenhagen:2005ga,Blumenhagen:2005pm}, 
    \begin{align}
        M_{ai} =
        \left\{
        \begin{array}{ll}
         2\pi {\rm tr}(T_a^2)m_a^i, &(i=1,2,\cdots, h^{1,1})
         \\
         \frac{1}{6}{\rm tr}(T_aT_bT_cT_d)\sum_{jkl}d_{jkl}m_b^jm_c^km_d^l +\frac{1}{24}{\rm tr}(T_a^2)\sum_jm_a^jc_{2,j}, &(i=0)
        \end{array}        
        \right.
        .
    \end{align}
    Hence, when $U(1)_Y$ gauge boson is defined as a linear combination of multiple $U(1)$s,
    \begin{align}
        U(1)_Y =\sum_a f_a U(1)_a,
    \end{align}
    the coefficient $f_a$ is constrained to prohibit the couplings 
    between the string axions and $U(1)_Y$ gauge boson,
    \begin{align}
        \sum_a M_{ai}f_a=0,
        \label{eq:U(1)Ymassless}
    \end{align}
    with $i=0,1,\cdots, h^{1,1}$, 
    otherwise $U(1)_Y$ gauge boson will become massive. 
     
    \item Generations of quarks and leptons
    
    Finally, we show the number of chiral zero-modes on the gauge background. Internal gauge fluxes as well as the curvature lead to the decomposition of the adjoint representation of $SO(32)$,
    \begin{align}
        496 \rightarrow \bigoplus_p (R_p, C_p),
    \end{align}
    where $R_p$ and $C_p$ denote the representations of the 4D gauge group $G$ and $W$, respectively. 
    From the  Hirzebruch-Riemann-Roch theorem, the net number of chiral zero-modes with $U(1)_a$ charge $Y_a$ is counted by the index
    \begin{align}
        \chi(L) &= 
        \int_{\cal M} \left[{\rm ch}_3(L) +\frac{1}{12}c_2(T{\cal M})\wedge c_1(L) \right]
        \nonumber\\
        &= \frac{1}{6}\sum_{i,j,k,a,b,c}d_{ijk}m_a^im_b^jm_c^k Y_aY_bY_c +\frac{1}{12}\sum_{i,a}c_{2,i}m_a^iY_a,
    \end{align}
    with $L=\otimes_{a}L_a^{Y_a}$. When we include the contribution from the freely-acting discrete symmetry group of CY ($\Gamma$), the index is divided by its order $|\Gamma|$, namely
    \begin{align}
        \chi(L) = -n|\Gamma|,
    \label{eq:Gen}
    \end{align}
    where $n$ represents the net number of chiral zero-modes regarded as the generations of the quarks, leptons/Higgs as well as the exotic particles.\footnote{For the freely-acting discrete symmetry group of CICY, see, \cite{Braun:2010vc}.} To prohibit chiral exotics in the low-energy effective action, we impose $\chi_{\rm exotic}=0$ for exotic particles. 
    To clarify the difference between 3-generation models and $n$-generation models in the autoencoder, we search for the $n$-generation models as discussed in detail in the next section. 
    \end{enumerate}

\subsection{$n$-generation models}

For definiteness, we focus on the decomposition of $SO(32)$ 
by the existence of multiple line bundles following Ref.~\cite{Otsuka:2018rki},
\begin{align}
    SO(32)&\rightarrow SO(16)\times SO(16)_H
    \nonumber\\
    &\rightarrow SO(6)\times SO(4)\times SO(2)^3\times SO(16)_H
    \nonumber\\
    &\rightarrow SU(3)_C\times SU(2)_L\times \Pi_{a=1}^5 U(1)_a\times SO(16)_H,  
\label{eq:Gdec}
\end{align}
where the internal gauge fluxes are inserted in the following $U(1)_a$ generators,
\begin{align}
    T_1 &={\rm diag}(1,1,1,0,0,0,0,0,0,\cdots, 0),
    \nonumber\\
    T_2 &={\rm diag}(0,0,0,1,1,0,0,0,0,\cdots, 0),
    \nonumber\\
    T_3 &={\rm diag}(0,0,0,0,0,1,0,0,0,\cdots, 0),
    \nonumber\\
    T_4 &={\rm diag}(0,0,0,0,0,0,1,0,0,\cdots, 0),
    \nonumber\\
    T_5 &={\rm diag}(0,0,0,0,0,0,0,1,0,\cdots, 0),
\end{align}
in the basis of Cartan directions of $SO(32)$ $H_r$ with $r=1,2,\cdots,16$ and $SO(32)$ roots are chosen as $(\underline{\pm 1, \pm 1, 0,\cdots, 0})$ under $H_r$. Here, the underline represents the possible permutation. 
Note that $U(1)_Y$ is a linear combination of the above five $U(1)_a$, $U(1)_Y=\sum_{a=1}^5 f_a U(1)_a$ and $SO(16)_H$ is regarded as a hidden gauge group. 
According to the gauge decomposition $SO(32)\rightarrow SO(16)\times SO(16)_H$, the adjoint representation of $SO(32)$ decomposes as
\begin{align}
496\rightarrow (120,1)\oplus (16, 16_H)\oplus (1, 120_H),
\end{align}
from which there exist the exotic particles in the $(16, 16_H)\oplus (1, 120_H)$,
\begin{align}
&(3, 1,16_H)_{1,0,0,0,0} \oplus (1, 2,16_H)_{0,1,0,0,0}  \oplus (1, 1,16_H)_{0,0,1,0,0} \oplus \nonumber\\
&(1, 1,16_H)_{0,0,0,1,0} \oplus (1, 1,16_H)_{0,0,0,0,1} \oplus (1, 1,120_H)_{0,0, 0,0,0}.
\end{align}
The subscript indices correspond to the $U(1)_a$ charges $Y_a$. 
On the other hand, the adjoint representation of $SO(16)$ includes the candidates of SM particles:
\begin{align}
\mathfrak{q} :~& (3, 2)_{-1,\pm 1,0,0,0}
,\nonumber\\
\mathfrak{l} :~& (1, 2)_{0,\pm 1, \pm 1,0,0} \oplus (1, 2)_{0,\pm 1, \mp 1,0,0} \oplus (1, 2)_{0,\pm 1, 0, \pm 1,0} \oplus (1, 2)_{0,\pm 1, 0, \mp 1,0} \oplus\nonumber\\ & (1, 2)_{0,\pm 1,0, 0, \pm 1} \oplus (1, 2)_{0,\pm 1,0, 0, \mp 1},\nonumber\\
\mathfrak{u}^c :~& (\bar{3}, 1)_{-2,0, 0,0,0} \oplus (\bar{3}, 1)_{1,0,\pm 1,0,0}\oplus (\bar{3}, 1)_{1,0,0,\pm 1,0}\oplus (\bar{3}, 1)_{1,0,0,0,\pm 1}
,\nonumber\\
\mathfrak{e}^c:~&(1, 1)_{0,\pm 2,0,0,0} \oplus (1, 1)_{0,0,\pm 1, \pm 1,0} \oplus (1, 1)_{0,0,\pm 1,\mp 1,0} \oplus (1, 1)_{0,0,\pm 1, 0, \pm 1}\oplus \nonumber\\
& (1, 1)_{0,0,\pm 1,0,\mp 1} \oplus (1, 1)_{0,0,0,\pm 1, \pm 1}\oplus (1, 1)_{0,0, 0,\pm 1,\mp 1}
.
\end{align}

Then, we define the quarks and lepton/Higgs in $\{\mathfrak{q},\mathfrak{l},\mathfrak{u}^c,\mathfrak{e}^c\}$ 
such that they have a proper hypercharge, 
\begin{align}\label{hypercharge constraint}
Q &= \left\{ q_x\in\mathfrak{q} \biggl|\,\sum_a f_a Y_a(q_x)=\frac{1}{6}\right\},
\nonumber\\
L &= \left\{ l_x\in\mathfrak{l} \biggl|\,\sum_a f_a Y_a(l_x)=-\frac{1}{2}\right\},
\nonumber\\
U^c &= \left\{ u_x^c\in\mathfrak{u}^c \biggl|\,\sum_a f_a Y_a(u_x^c)=-\frac{2}{3}\right\},
\nonumber\\
D^c &= \left\{ d_x^c\in\mathfrak{u}^c \biggl|\,\sum_a f_a Y_a(d_x^c)=\frac{1}{3}\right\},
\nonumber\\
E^c &= \left\{ e_x^c\in\mathfrak{e}^c \biggl|\,\sum_a f_a Y_a(e_x^c)=1\right\},
\end{align}
where it is noted that $Y_a(\phi)$ stands for the $U(1)_a$ 
charge of the field $\phi$. 
When the particles in $\{\mathfrak{q},\mathfrak{l},\mathfrak{u}^c,\mathfrak{e}^c\}$ 
satisfy these conditions, they are identified with the quarks and leptons/Higgs, otherwise the others do not belong to the spectra in the SM, regarded as the exotic particles. From the view point of the representations of the SM gauge group, we cannot distinguish between Higgsino fields and the charged leptons, but it is distinguishable when we clarify the $SO(32)$ gauge invariant Yukawa couplings among elementary particles, such as $QH_uU^c,QH_dD^c,LH_dE^c$. 

In the neural network implemented in the next section, 
we impose the following phenomenological constraints in 
addition to the consistency conditions in Section~\ref{sec:2_1}:\footnote{Note that the hypercharge masslessness condition (\ref{eq:U(1)Ymassless}) is simplified as ${\rm tr}(T_a^2)m_a^if_a=0$ in the gauge decomposition (\ref{eq:Gdec}).}
\begin{eqnarray}
n_Q=n_L=n_{U^c}=n_{D^c}=n_{E^c}=n, \qquad
n_\phi=0\qquad (\forall \phi\in {\rm exotics}),
\end{eqnarray}
with
\begin{eqnarray}
n_Q\equiv \sum_{\mathfrak{q}\in Q}n_\mathfrak{q},\quad n_L\equiv \sum_{\mathfrak{l}\in L}n_\mathfrak{l},\quad n_{U^c}\equiv \sum_{\mathfrak{u}^c\in U^c}n_{\mathfrak{u}^c},\quad n_{D^c}\equiv \sum_{\mathfrak{u}^c\in D^c}n_{\mathfrak{u}^c},\quad n_{E^c}\equiv \sum_{\mathfrak{e}^c\in E^c}n_{\mathfrak{e}^c},
\end{eqnarray}
where each $n_\ast$ is evaluated by employing Eq.~(\ref{eq:Gen}).

\section{Classification methods for $SO(32)$ line bundle models}
\label{sec:3}

In this section, we show the detailed method to apply the machine learning 
technique to the heterotic line bundle models satisfying the several 
conditions discussed in the previous section. Especially, we restrict ourselves to CICY threefolds with Hodge number $h^{1,1}\leq 5$. 
There exist 5, 36, 155, 425 and 856 CICY threefolds with $h^{1,1}=1,2,3,4$ and $5$ respectively. 
Before going to the detailed description, we outline the processing flow given in the following four key steps:
\begin{enumerate}
    \item Make dataset of $n$-generation models on the large number of CICYs, satisfying the conditions in Section \ref{sec:2}.
    \item Reduce the dimension of input parameters to the 2D charts by the autoencoder.
    \item Classify the model data based on the results of dimension reduction using k-means++ algorithm. Then, calculate the percentage of 3-generation models for each cluster in the 2D space and decide ``3-generation island".
    \item Find the difference between the 3-generation island and other region. This corresponds to the feature of 3-generation model.
\end{enumerate}

In the following subsections, we give a detailed description of each step. 

\subsection{Collect data}

The first step of our method is to obtain a dataset of line bundle models 
satisfying constraints discussed in Section \ref{sec:2}. It is known that solving the constraints is mathematically difficult because they have many integer variables in the equations 
called Diophantine equations. It was proved that there is no general method to solve 
this equation even though all of the constraints are polynomial\cite{Diophantine}. 
Then we adopt a brute force approach as detailed in \algref{alg:brute force search} in which we employ the following simplification \cite{Otsuka:2018rki}. 
Since there is no summation over $i$ in K-theory condition (\ref{simplified K-theory condition}) and the 
hypercharge masslessness condition (\ref{eq:U(1)Ymassless}), both are rewritten as
\begin{align}
    m_\alpha^i = \sum_{A=3,4,5} K_{\alpha A}m_A^i\quad (\alpha=1,2),
\end{align}
with
\begin{align}
    K_{1A}(T_a,f_a) &=\frac{\tr (T_2)\tr (T_A^2)f_A-\tr (T_A)\tr(T_2^2)f_2}{\tr(T_1)\tr(T_2^2)f_2-\tr (T_2)\tr (T_1^2)f_1},\nonumber\\
    K_{2A}(T_a,f_a) &=\frac{\tr (T_1)\tr (T_A^2)f_A-\tr (T_A)\tr(T_1^2)f_1}{\tr(T_2)\tr(T_1^2)f_1-\tr (T_1)\tr (T_2^2)f_2}.
\end{align}
When the K-theory condition is given by (\ref{K-theory condition}), only the hypercharge masslessness condition boils down to
\begin{align}
    m_1^i = \sum_{A=2,3,4,5} K'_{A}(T_a,f_a)m_A^i,
\end{align}
with
\begin{align}
    K'_A(T_a,f_a)&=-\frac{\tr (T_A^2)f_A}{\tr (T_1^2) f_1}.
\end{align}

\begin{algorithm}
\caption{Brute force search}
\label{alg:brute force search}
\begin{algorithmic}[1]
\State{Decide search region as $h^{1,1}\leq h^{1,1}_*$ and $-m_*\leq m_a^i \leq m_*$. Give the upper bound for the number of times the random attack $R_{\rm max}^1$ and $R_{\rm max}^2$.}
\State{Calculate the topological data of CICY $d_{ijk}, c_{2,i}$ within $h^{1,1}\leq h^{1,1}_*$.}
\If{K-theory condition is given by \eqref{simplified K-theory condition}}\label{state:begin_search}
\State{Since there is no summation over $i$ in \eqref{simplified K-theory condition} and \eqref{eq:U(1)Ymassless}, a part of $m_a^i = \mu_a$ is determined by
    \begin{equation}
        \mu_\alpha=\sum_{A=3,4,5}K_{\alpha A}(T_a,f_a)\mu_A\quad(\alpha=1,2).
    \end{equation}}
\State{Choose $-m_*\leq \mu_A\leq m_*$ at random and confirm $-m_*\leq \mu_\alpha \leq m_*$.}
\Else
\State{From \eqref{eq:U(1)Ymassless}, only $m_a^1 = \mu_1$ is determined by
    \begin{equation}
        \mu_1=\sum_{A=2,3,4,5}K'_{A}(T_a,f_a)\mu_A.
    \end{equation}}
    \State{Choose $-m_*\leq \mu_A\leq m_*$ at random and confirm $-m_*\leq \mu_\alpha \leq m_*$ and \eqref{K-theory condition}.}
\EndIf
\State{Then we obtain a flux list $\{\vec\mu\}=\{\mu_1,\mu_2,\mu_3,\mu_4,\mu_5\}$ with the number $N_{\mu}$.}
\If{$R_{\rm max}^1\leq\, _{N_\mu}P_{h^{1,1}}\small{(=N_\mu!/(N_\mu-h^{1,1})!)}$}
\For{$R_{\rm max}^1$ times}
\State{Construct $m_a^i$ 
from $\{\vec\mu\}$ at random.}
\State{Find models satisfying \eqref{eq:NS5 condition} and \eqref{eq:Gen}.}
\EndFor
\Else
\For{all possible patterns of constructing $m_a^i$ from $\{\vec\mu\}$}
\State{Find models satisfying \eqref{eq:NS5 condition} and \eqref{eq:Gen}.}
\EndFor
\EndIf
\State{As a result, we obtain $n$-generation models satisfying the phenomenological and theoretical conditions.}\label{state:end_search}
\State{However, the obtained models have typically 0-generation of quarks and leptons. 
    Hence, let us extract $f_a$ leading to nonzero-generation models from the possible $f_a$ list.}
\State{Repeat step \ref{state:begin_search} to step \ref{state:end_search} for the specific $f_a$ with replacing the 
random attack $R_{\rm max}^1$ by $R_{\rm max}^2$.}
\State{Finally we obtain many $n\neq 0$-generation models.}
\end{algorithmic}
\end{algorithm}

We carry out the above brute force approach for three times named as search (I), (II) and (III) with different parameter region shown in Table \ref{tab:search_region}. The obtained number of $n$-generation models for each search is summarized in Table. \ref{tab:numofmodels}. When $h^{1,1}=2$ in the search (I) and (II), it is difficult to realize the $n$-generation models. Then, we concentrate on the other 7 patterns in the  search (I)-(III).
\begin{table}[H]
    \label{search region}
    \centering
    \begin{tabular}{|c|c|c|c|c|c|}\hline
    Search&$h^{1,1}_*$&$m_*$&K-theory condition&$R_{\rm max}^1$&$R_{\rm max}^2$\\ \hline
    (I) & 5& 2& \eqref{simplified K-theory condition}&$10^4$&$5\times 10^5$\\
    (II) & 4& 3& \eqref{simplified K-theory condition}&$10^3$&$3\times 10^6$\\
    (III) & 3& 2& \eqref{K-theory condition}&$10^4$&$10^6$\\ \hline
    \end{tabular}
    \caption{Maximal values of Hodge number $h^{1,1}_\ast$ and the flux $m_\ast$, K-theory condition and the number of random attacks $R_{\rm max}^{1,2}$ for each search}.
\label{tab:search_region}
\end{table}

\begin{table}[H]
    \centering
    \begin{tabular}{|c|c|c|c|c|}\hline
    Search&$h^{1,1}$&\# of $n\neq 0$ models&\# of $n=3$ models& Percentage of $n=3$ models (\%)\\ \hline
    &2&24& 0&0.00\\
    (I) & 3& 18072&463&2.56\\
    &4&15622&271&1.73\\
    &5&5843&128&2.19 \\ \hline
    &2&120&0&0.00\\
    (II) & 3& 12293&178&1.45\\
    &4& 5448&62&1.14\\ \hline
    (III)&2&5664& 768&13.56\\
    & 3& 23826&1152&4.84\\ \hline
    \end{tabular}
    \caption{The number of $n\neq 0$-generation models and 3-generation models for each search with a specific $h^{1,1}$.}
\label{tab:numofmodels}   
\end{table}

\subsection{Autoencoder}
After collecting the data of $n$-generation line bundle models, 
we perform the autoencoder known as a kind of multi-layer perceptrons (MLP). 
The advantage of the autoencoder is to reduce the higher-dimensional parameter 
space of the input data to the compressed data in the 2D charts and at the same time, to extract characteristic features of the data without giving any information how to extract the data. 
The fundamental component of the MLP is called a perceptron, which transforms a $N_0$-dimensional vector $\Vec{x}_0$ into a number $x_1$
\begin{equation}
    x_1=h(\vec{w}'\cdot\vec{x}+b'),
\end{equation}
where $h$ is a generally non-linear function typically chosen as a sigmoid function or ReLU function, $\vec{w}'$ is a $N_1$-dimensional vector called weight and the number $b'$ represents a bias. A layer consists of multiple perceptrons. When the layer consists of $N_1$ perceptrons, they transform $N_0$-dimensional vector into $N_1$-dimensional vector, and weight and bias become a $N_0\times N_1$ matrix and $N_1$ dimensional vector respectively.

\begin{figure}
    \centering
    \includegraphics[width=14cm]{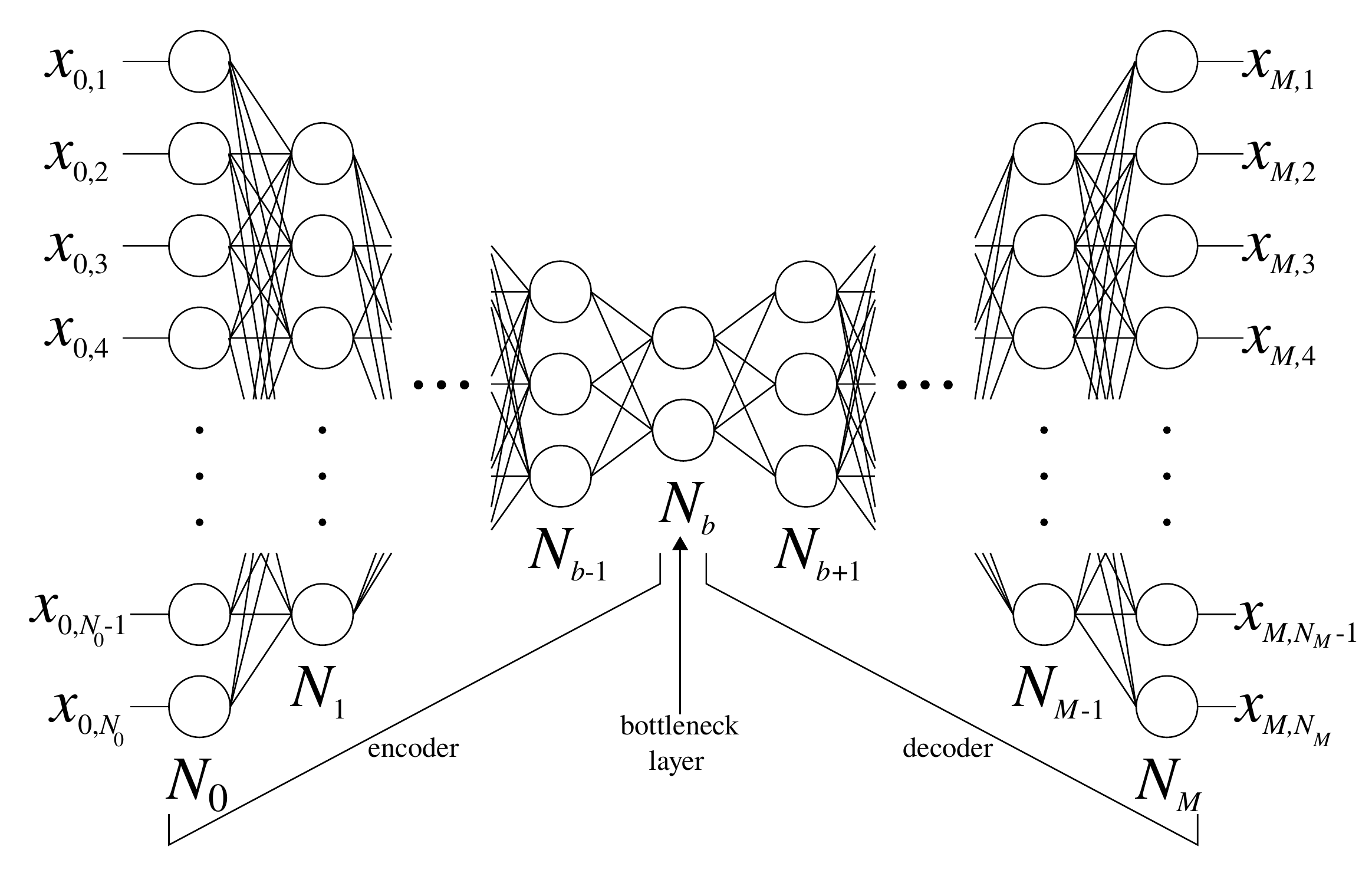}
    \caption{Schematic diagram of an autoencoder. Each circle represents one perceptron.}
    \label{fig:AE}
\end{figure}

Let us suppose that the input data is the $N_0$-dimensional vector and the output is $N_M$-dimensional vector. Then, the $n$-th layer in MLP has  $N_{n-1}$-dimensional input vector $\vec{x}_{n-1}=(x_{n-1,1},\cdots,x_{n-1,N_{n-1}})$ and $N_n$-dimensional output vector $\vec{x}_n=(x_{n,1},\cdots,x_{n,N_n})$. These two are related by
\begin{equation}
    x_{n,i}=h_n(w^n_{ij}x_{n-1,j}+b^n_i),
\end{equation}
where the weight $w^n$ and the bias $b^n$ are described by $N_n\times N_{n-1}$ matrix 
and $N_n$ components, respectively. 
In this way, MLP is constructed by connecting $M$ layers in series, as drawn in \figref{AE}. 
In the context of MLP, learning corresponds to tune the weights and biases to minimize an error function $E(\vec{x}_M,\vec{y})$ representing the difference between training data $\vec{y}$ and outputs of MLP $\vec{x}_M$.

The autoencoder consists of the MLP with $N_0>N_1>\cdots > N_b$ and $N_b < N_{b+1} < \cdots < N_M=N_0$ ($b=(M-1)/2$) as shown in \figref{AE}, in which the first and latter half are called encoder and  decoder, respectively. 
Here, we denote the training data by $\vec{x}_0$ and design $\vec{x}_0$ such that 
the outputs resemble the inputs as closely as possible.
After learning, $\vec{x}_{b}$ has lower dimension than $\vec{x}_0$ but it is possible to construct $\vec{x}_M=\vec{x}_0$. It indicates that all information (features) of inputs is compressed into the outputs of $b$-th layer.

The application of the autoencoder to our system is described as follows. 
We take the input vector as $(|\Gamma|, d_{ijk}, c_{2,i}, m_a^i)$ deciding the line bundle models. When $h^{1,1}\leq 5$, dimension of this vector is not more than 161 dimension, hence we take $N_0=N_M=161$. Note that the meaningless parameters such as $d_{145}$ for CICY threefolds with $h^{1,1}=3$ are filled by 0 and this procedure does not affect any constraints we consider. 
In addition, we take $N_b=2$ to compare our results with the heterotic orbifold 
results \cite{Mutter:2018sra} and to 
visualize the result easily. 
After trial and error, we arrive at the expression for 7 layers in the encoder and decoder with the following dimensions,
\begin{equation}
    N_n=(161,80.40,20.10,5,3,2,3,5,10,20,40,80,161).
\end{equation}
The activation functions are chosen as sigmoid functions for $h_1,\cdots,h_b$ and 
identical maps for $h_0,h_{b+1},\cdots,h_{M}$, respectively and the error function 
is given by
\begin{equation}
    E(\vec{x}_0,\vec{x}_M)=|\vec{x}_M-\vec{x}_0|^2.
\end{equation}
The learning method is followed by Adam-Optimizer in TensorFlow\cite{Abadi:2016kic}. To avoid becoming trapped in a local minima of the error function, we first decompose the autoencoder into 
partial autoencoders with three-layer $(N_{q-1},N_q,N_{M+1-q})\:(q=1,\cdots,7)$ and after that the whole autoencoder is learned to minimize the error function.
The learning is repeated (40000, 20000, 20000, 20000, 18000, 16000, 14000) times for each partial autoencoder and 20000 times for the whole autoencoder. 
Since the two-dimensional scatter plots of the bottleneck layer have a 
cluster structure as demonstrated later, we apply the clustering method to 
the result of autoencoders.

\subsection{K-means++ clustering}
To classify the compressed information in the bottleneck layer, 
we adopt the famous k-means++ method which has advantages that the algorithm itself is simple and computational cost is not significant. 
We employ KMeans class in scikit-learn for k-means++ clustering\cite{sklearn}. 

K-means++ classifies a given data $\mathcal{D}$ with distance $d(\cdot,\cdot)$ into $N_{\rm cl}$ clusters as explained in \algref{alg:k-means++}. In our case, $\mathcal{D}$ 
corresponds to the encoded vectors $\vec{x}_b$ and we take the distance as Euclidean  norm.

\begin{algorithm}
\caption{K-means++ clustering}
\label{alg:k-means++}
\begin{algorithmic}[1]
\State{Give $\mathcal{D}$ and $N_{\rm cl}$.}
\State{Pick up randomly $x_1\in\mathcal{D}$.}
\For{$i=2,\cdots,N_{\rm cl}$,}
    Pick up $x_i$ from $\mathcal{D}$ with probability 
    \begin{equation}
        \frac{\min_{X=x_1,\cdots,x_{i-1}}d(X,x_i)^2}{\sum_{x\in\mathcal{D}}\min_{X=x_1,,\cdots,x_{i-1}}d(X,x_i)^2}
    \end{equation}
\EndFor
\State{Define clusters $C_l=\{x\in\mathcal{D}|d(x,x_l) < d(x,x_m)\:\forall m\neq l\}\:(l=1,\cdots,N_{\rm cl})$.}
\State{Find the centroid of $C_l$ named as $g_l$.}\label{findcl}
\State{Re-classify as $C_l=\{x\in\mathcal{D}|d(x,g_l) < d(x,g_m)\:\forall m\neq l\}\:(l=1,\cdots,N_{\rm cl})$.}\label{reclassify}
\State{Repeat the steps \ref{findcl} and \ref{reclassify} until $\{g_l\}$ converge.}
\State{Finally, we result in $C_l=\{x\in\mathcal{D}|d(x,g_l) < d(x,g_m)\:\forall m\neq l\}\:(l=1,\cdots,N_{\rm cl})$.}
\end{algorithmic}
\end{algorithm}

To decide an appropriate $N_{\rm cl}$, we employ so called elbow method explained in \algref{alg:elbow method} from which a critical value of $N_{\rm cl}$ ($N_{\rm cl}^*$) is estimated. Then, we tune the number of clusters around $N_{\rm cl}^*$ by eye-estimation in order to extract the 3-generation structure significantly.

\begin{algorithm}[H]
\caption{Elbow method}
\label{alg:elbow method}
\begin{algorithmic}[1]
\State{Give the maximal value of $N_{\rm cl}$ named as $N_{\rm cl}^{\rm max}$ and number of trials $N_{\rm trial}$.}
\For{$N_{\rm cl}=1,\cdots,N_{\rm cl}^{\rm max}$}
\State{Perform k-means++ clustering $N_{\rm trial}$ times and calculate ``distortion" for each trial
\begin{equation}
    D(\mathcal{D},N_{\rm cl},\{C_l\})\overset{\rm{def}}{=}\sum_{l=1,\cdots,N_{\rm cl}}\sum_{x\in C_l}|g_l-x|^2.
\end{equation}
}\label{trials}
\State{Define 
\begin{equation}
    D(\mathcal{D},N_{\rm cl})=\min_{\rm trials\,in\,Step\, \ref{trials}}D(\mathcal{D},N_{\rm cl},\{C_l\}).
\end{equation}
}
\EndFor
\State{For some $N_{\rm cl}^*$, the distortion is saturated i.e.
\begin{equation}
    \frac{D(\mathcal{D},N_{\rm cl}+1)}{D(\mathcal{D},N_{\rm cl})}\sim 1 \quad \forall N_{\rm cl}\geq N_{\rm cl}^*.
\end{equation}
}
\State{This $N_{\rm cl}^*$ is considered a suitable number of clusters 
indicated from the elbow method.}
\end{algorithmic}
\end{algorithm}

\subsection{Statistical analysis}
To find the factor of differences between the 3-generation island and other region, we introduce the KL divergence KL$(\rho_1,\rho_2)$ which represents the distance between two distributions $\rho_1$ and $\rho_2$. It is defined by
\begin{equation}
    {\rm KL}(\rho_1,\rho_2)=\sum_m \rho_1(m)\log\frac{\rho_1(m)}{\rho_2(m)},
\end{equation}
where $\rho(m)$ denotes a probability of taking $m$ under $\rho$. 
Note that KL$(\rho_1,\rho_2)=0$ under $\rho_1=\rho_2$.

In our case, $\rho_1$ and $\rho_2$ stand for the distributions of an input parameter $X\in(d_{ijk},m_a^i,c_{2,i},|\Gamma|)$ of the 3-generation island and all region, respectively. 
In the following, we define KL$(X)\overset{\mathrm{def}}{=}{\rm KL}(\rho_1,\rho_2)$. 
Note that $X$ with small KL$(X)$ does not contribute to the identification of the 3-generation island, 
whereas $X$ with large KL divergence plays a crucial role in distinguishing between the 3-generation island and other region.

\section{Results}
\label{sec:4}

In this section, we summarize the results by implementing the autoencoder and k-means++ clustering in $SO(32)$ heterotic line bundle models. 
After discussing the search (I) in detail in Section~\ref{sec:4_1}, 
we show the other searches in Section~\ref{sec:4_2}. 
Finally, we count the number of Higgs pairs which are vector-like under 
the SM gauge group, but chiral with respect to other extra $U(1)$s 
by checking the Yukawa couplings of quarks and leptons.

\subsection{Search (I) with $h^{1,1}=3$ and $N_{\rm cl}=26$}
\label{sec:4_1}

\begin{figure}[H]
\begin{minipage}{0.5\hsize}
    \centering
    \setlength{\captionmargin}{15pt}
    \includegraphics[width=9cm]{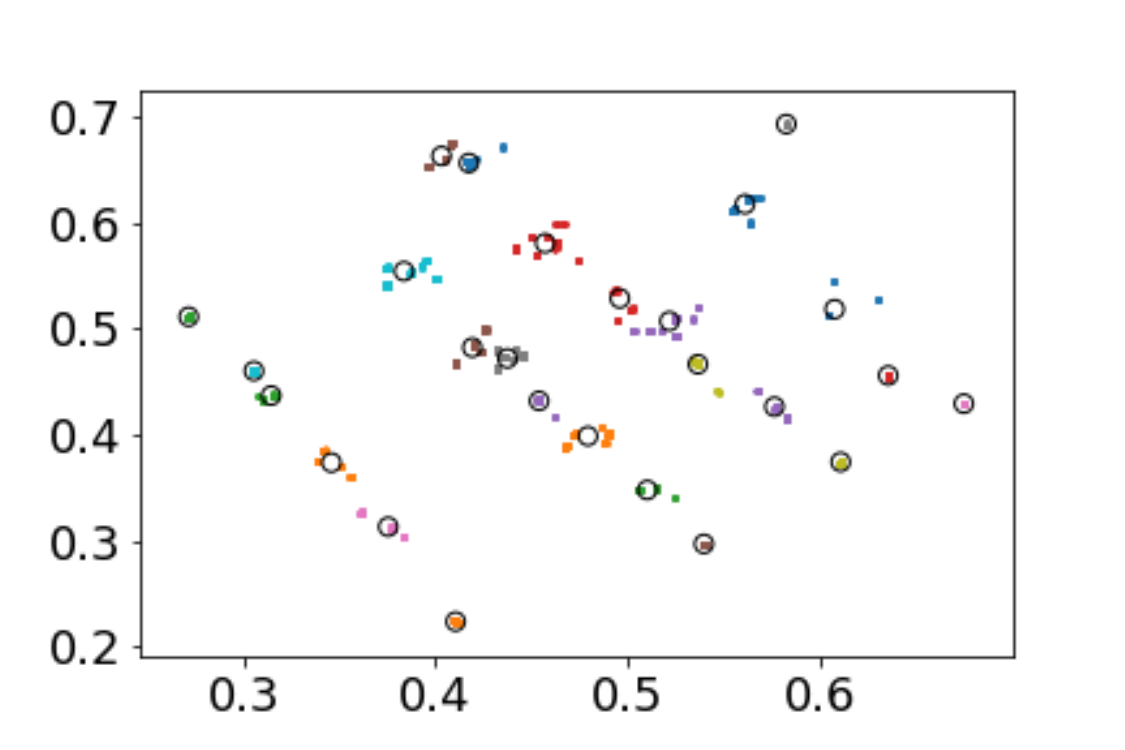}
    \caption{Result of the autoencoder and k-means++ clustering, where the horizontal and vertical axes represent the first and second components of $\vec{x}_b$, respectively. Circles correspond to centroids of 
    colored clusters.\\}
    \label{fig:clustering}
\end{minipage}
\begin{minipage}{0.5\hsize}
    \centering
    \setlength{\captionmargin}{15pt}
    \includegraphics[width=9cm]{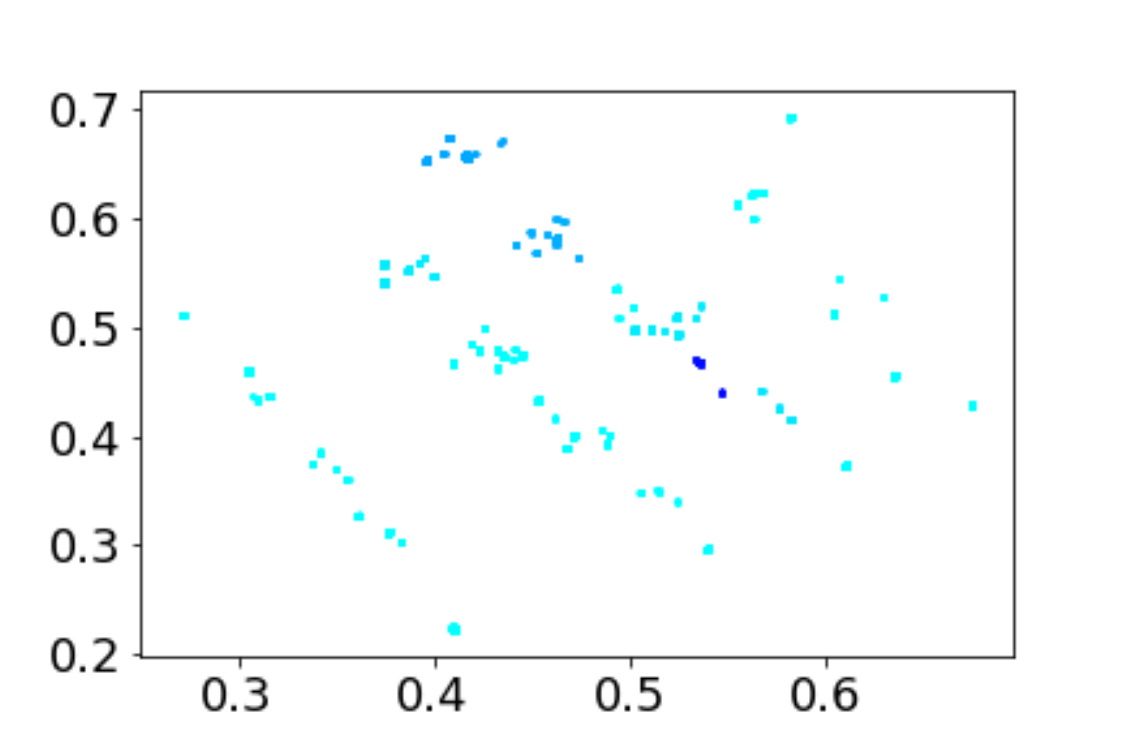}
    \caption{Ratio of 3-generation models to $n\neq0$-generation models in each cluster, 
    where the higher the ratio, the darker the color is. The horizontal and vertical axes represent the first and second components of $\vec{x}_b$, respectively.}
    \label{fig:3genisland}
\end{minipage}
\end{figure}

First, we discuss the case of search (I) with $h^{1,1}=3$ and $N_{\rm cl}=26$ as a concrete example. 
\figref{clustering} shows the result of k-means++ clustering of $\vec{x}_b$ at the bottleneck layer  and black circles correspond to centroids of each colored cluster.
\figref{3genisland} represents the ratio between $n=3$ models and $n\neq0$-generation models in each cluster. The density of 3-generation models in the deep blue region is higher than the other region. The cluster located around (0.54,0.47) in \figref{3genisland} has the highest ratio ($\cong 19.15\%$) among total 26 clusters and then this cluster is identified with the 3-generation island. 
Recalling that this 3-generation island contains only 2.08\% of the whole line bundle models we consider, it is easy to find the 3-generation models by focusing on this fertile island. Such a phenomena is also discussed in the heterotic $\mathbb{Z}_6$-II orbifold landscape~\cite{Mutter:2018sra}. 
It is remarkable that 19 clusters in all 26 clusters do not have 3-generation model. 
In this respect, we argue that our clustering procedure extracts features of $n=3$ models.

\begin{figure}[H]
    \centering
    \includegraphics[width=17cm]{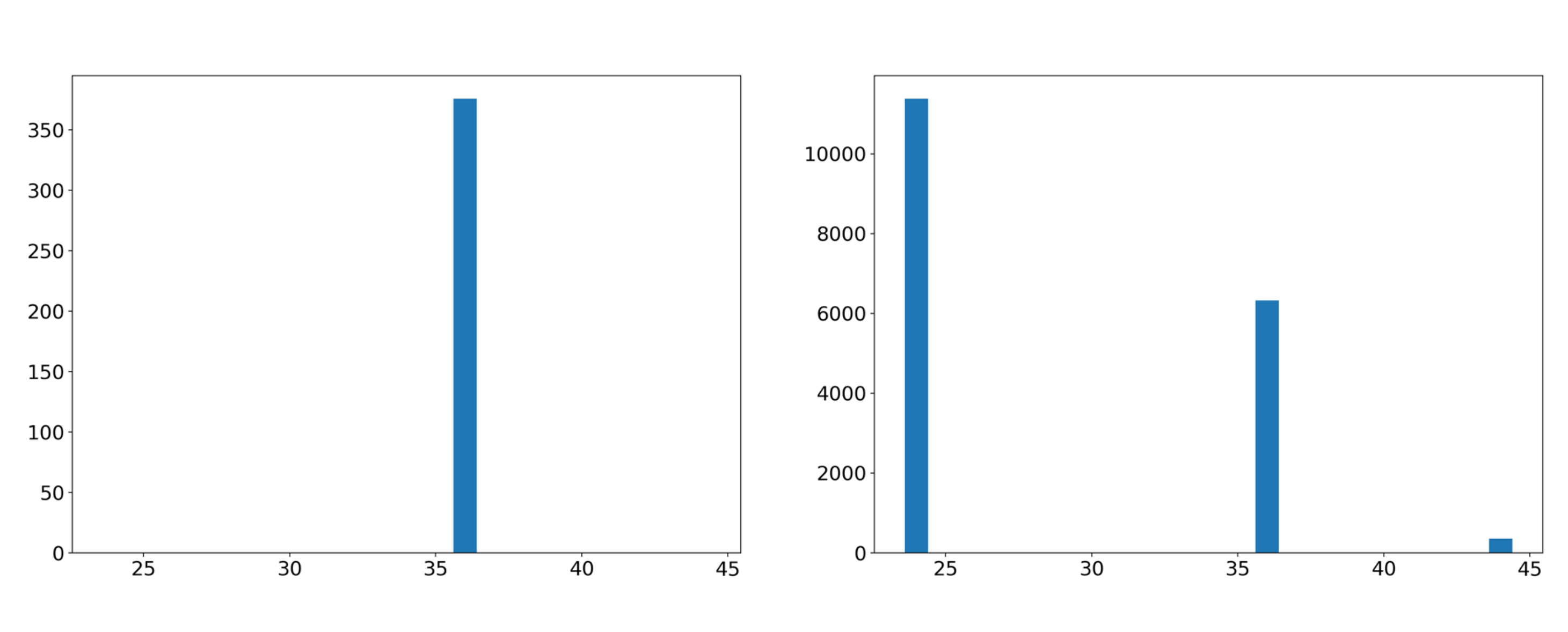}
    \caption{The histograms of $c_{2,1}$ for the 3-generation island in the left panel and all region in the right panel, where the horizontal and vertical axes represent the value of $c_{2,1}$ and the number of models respectively.}
    \label{fig:c21}
\end{figure}

\begin{figure}[H]
    \centering
    \includegraphics[width=17cm]{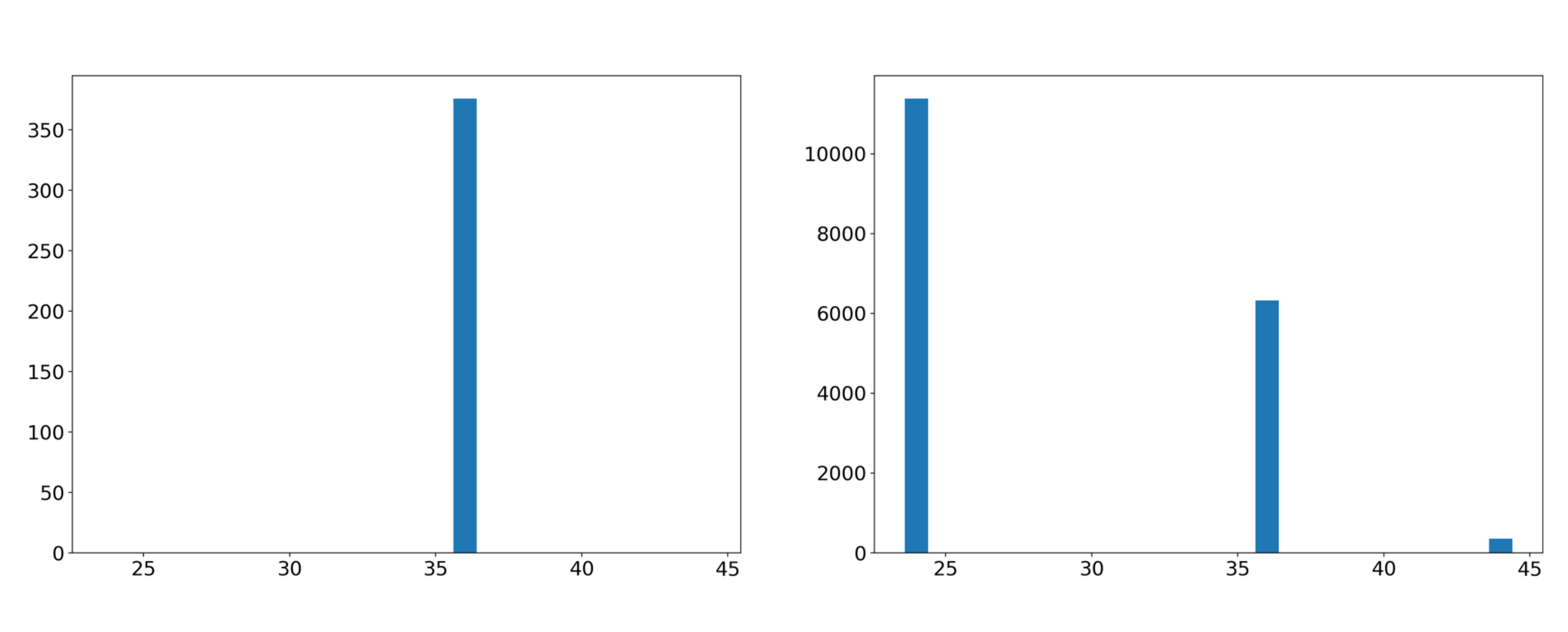}
    \caption{The histograms of $c_{2,2}$ for the 3-generation island in the left panel and all region in the right panel. The axes are the same as \figref{c21}.}
    \label{fig:c22}
\end{figure}

\begin{figure}[H]
    \centering
    \includegraphics[width=17cm]{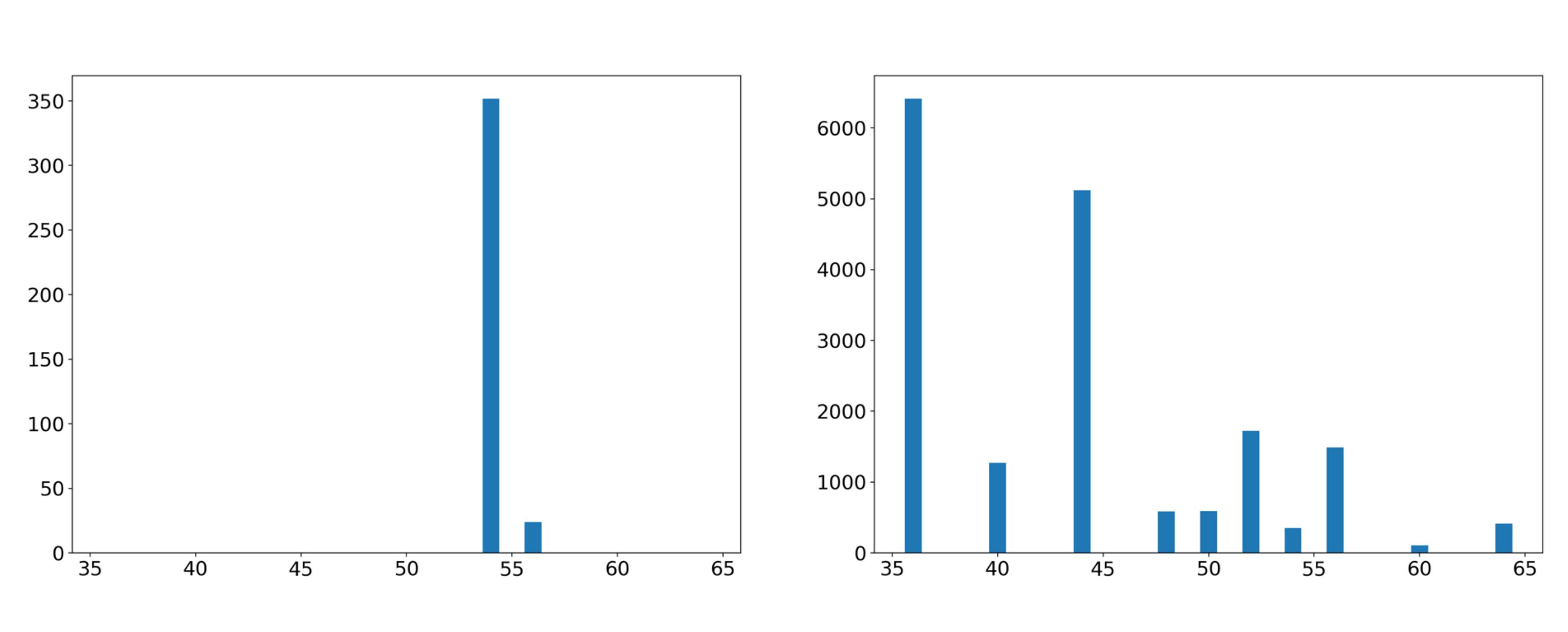}
    \caption{The histograms of $c_{2,3}$ for the 3-generation island in the left panel and all region in the right panel. The axes are the same as \figref{c21}.}
    \label{fig:c23}
\end{figure}

In Table \ref{tab:KL}, we list the KL$(X)$ for all $X$ where $X=d_{ijk}$ are shown only $i\leq j\leq k$ because of the permutation symmetry of the indices. As mentioned above, $X$ with large KL$(X)$ captures an 
information of the 3-generation island. We find that 3-generation models have a strong correlation with the topological data of CY rather than flux parameters. Especially,  characteristic values of $c_{2,i}$, in particular 36 or 54, are selected as shown in Figs. \ref{fig:c21}, \ref{fig:c22} and \ref{fig:c23}. We conclude that 3-generation models prefer $c_{2,i}\in18\mathbb{Z}$, although there is no bias in  $c_{2,i}$ we employed. Small KL$(d_{111})$ in Table \ref{tab:KL} indicates that almost all of CICYs with $h^{1,1}=3$ leading to $n$-generation models have the triple self-intersection number $d_{111}=0$. Other $d_{ijk}$ are typically 0, 3 or 9. 
Such specific values also represent a characteristic feature in the 3-generation island due to their large KLs, but they take different values for different $h^{1,1}$ and/or searches.
\begin{table}[H]
    \centering
    \begin{tabular}{|c|c|c|c|c|c|c|c|}\cline{1-2} \cline{4-5} \cline{7-8}
        $X$&KL$(X)$&&$X$&KL$(X)$&&$X$&KL$(X)$ \\ \cline{1-2} \cline{4-5} \cline{7-8}
$d_{333}$&3.6089&&$m_1^3$&0.5836&&$m_5^3$&0.1996\\ \cline{1-2} \cline{4-5} \cline{7-8}
$c_{2,3}$&3.6089&&$c_{2,2}$&0.5405&&$m_2^2$&0.1479\\ \cline{1-2} \cline{4-5} \cline{7-8}
$d_{133}$&3.4667&&$|\Gamma|$&0.5175&&$m_1^2$&0.1363\\ \cline{1-2} \cline{4-5} \cline{7-8}
$d_{233}$&3.3863&&$m_1^1$&0.3724&&$m_3^1$&0.0481\\ \cline{1-2} \cline{4-5} \cline{7-8}
$d_{123}$&2.7453&&$m_2^3$&0.3696&&$m_4^1$&0.0453\\ \cline{1-2} \cline{4-5} \cline{7-8}
$d_{113}$&1.1448&&$f_a$&0.3502&&$m_5^1$&0.0452\\ \cline{1-2} \cline{4-5} \cline{7-8}
$d_{112}$&1.1343&&$d_{222}$&0.2875&&$m_4^2$&0.0278\\ \cline{1-2} \cline{4-5} \cline{7-8}
$c_{2,1}$&1.0503&&$m_3^3$&0.2815&&$m_5^2$&0.0267\\ \cline{1-2} \cline{4-5} \cline{7-8}
$d_{122}$&1.0494&&$m_4^3$&0.2681&&$d_{111}$&0.0201\\ \cline{1-2} \cline{4-5} \cline{7-8}
$d_{223}$&1.0482&&$m_2^1$&0.2100&&$m_3^2$&0.0167\\ \cline{1-2} \cline{4-5} \cline{7-8}
    \end{tabular}
    \caption{KL divergence of each parameter.}
    \label{tab:KL}
\end{table}

\subsection{Other cases}
\label{sec:4_2}

We perform a similar analysis for other searches. In the case of search (III) with $h^{1,1}=2$, 
there are two islands with the same percentage of $n=3$ models and we define both of two as the 3-generation island. 
\begin{table}[H]
    \centering
    \begin{tabular}{|c|c|c|l|c|c|}\hline
        Search & $h^{1,1}$ & $N_{\rm cl}$ & \multicolumn{1}{|c|}{Favored $c_{2,i}$}&\# of $n\neq 0$ models& Percentage of $n=3$ models\\
        &&&&in the 3-generation island&in the 3-generation island(\%)\\ \hline
        &3&26&(36,36,54)&376&19.15\\
        (I)&4&30&(24,24,36,36)&1095&6.03\\
        &5&42&(24,36,36,36,36)&57&17.54\\ \hline
        (II)&3&40&(36,36,36)&235&24.26\\
        &4&30&(24,36,36,36)&93&4.30\\ \hline
        (III)&2&12&(36,36)&1536&25.00\\ 
        &3&27&(36,36,54)&960&20.21\\ \hline
    \end{tabular}
    \caption{Number of clusters $N_{\rm cl}$, \# of $n\neq 0$ models and percentage of $n=3$ models in the 3-generation island. Favored $c_{2,i}$ in the 3-generation island is listed.} 
    \label{tab:favored c2i}
\end{table}

From Table~\ref{tab:favored c2i} summarizing all our searches, 
particular values of the second Chern numbers of CICYs 
are favored in the class of 3-generation models, although there is no bias in the topological data of CICYs we employed. For the cases of $c_{2,i}=24$, their KL divergences are relatively smaller than others, indicating that these cases are not important for analysis of 3-generation models. 
For instance, in search (I) with $h^{1,1}=4$, KL$(c_{2,1})=0.026$ and KL$(c_{2,2})=0.231$ are suppressed compared 
with KL$(c_{2,3})=0.558$ and KL$(c_{2,4})=1.363$. 
From the above discussion, favored second Chern numbers in the 3-generation island are provided by $c_{2,i}\in18\mathbb{Z}$. 
Then we conclude that 3-generation models have a strong correlation with $c_{2,i}\in 18\mathbb{Z}$.

%4/19
%Here we comment on the 3-generation island in 161-dimensional parameter space.
Here we comment on whether the 3-generation 
%island is realized in the decoder described by (at most) 161-dimensional parameter spaces. 
island is realized by the decoder described in (at most) 161-dimensional parameter spaces. 
The region obtained by feeding points in the 3-generation island to the decoder is a two-dimensional plane embedded in the 161-dimensional parameter spaces because our activation functions of decoder are the identical map,
although the relations of parameters in the 161-dimensional spaces are non-linear. 
Then the decoder is not competent for specifying the 3-generation island in the 161-dimensional spaces.

Let us also comment on the geometrical interpretation of this second Chern number of CICY. 
The instanton number of the tangent bundle of CY $T{\cal M}$ with the curvature two-form $R$ is given by
\begin{equation}
    N_{\rm ins}^{T{\cal M}}=-\frac{1}{8\pi^2}\int_{\cal M}\tr R\wedge *R=-\frac{1}{8\pi^2}\int_{\cal M}\tr R\wedge R\wedge \Omega,
\end{equation}
where we employ the self-dual condition of the curvature two-form
\begin{equation}
    *R=\Omega\wedge R
\end{equation}
with $\Omega$ satisfying the condition $d\Omega =0$\cite{Kanno}.  
Recall that K$\mathrm{\ddot{a}}$hler form of CY manifolds is a closed form, 
it is possible to take $\Omega=w_i$, namely
\begin{equation}
    N_{{\rm ins},i}^{T{\cal M}}= -\frac{1}{8\pi^2}\int_{\cal M}\tr R\wedge R\wedge w_i=\int_{\cal M} c_2(T{\cal M})\wedge w_i = c_{2,i}
\end{equation}

It is interesting to ask why these specific instanton numbers are favored in a class of 
3-generation models. We hope to report on this relationship in the future. 

\subsection{Number of generations of Higgs}
\label{sec:4_3}

\begin{figure}[H]
\begin{minipage}{0.5\textwidth}
    \centering
    \includegraphics[width=7.2cm]{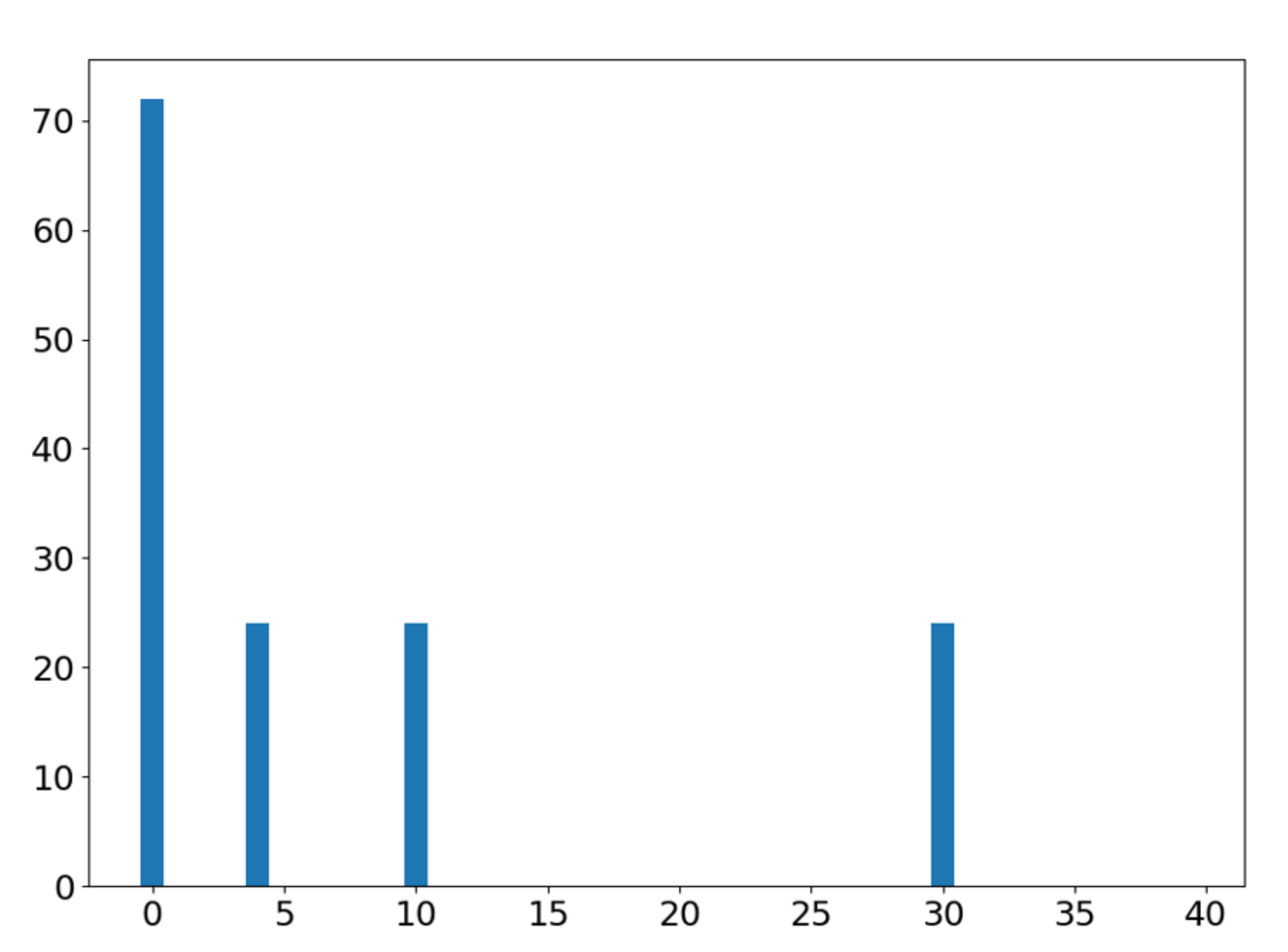}
    \setlength{\captionmargin}{10pt}
    \caption{\# of $n$-generation models w.r.t. $n_H$ for search (I) with $h^{1,1}=3$. The horizontal and vertical axes represent $n_H$ and the number of models, respectively.}
    \label{fig:Higgshist1}
\end{minipage}
\begin{minipage}{0.5\textwidth}
    \centering
    \includegraphics[width=7.2cm]{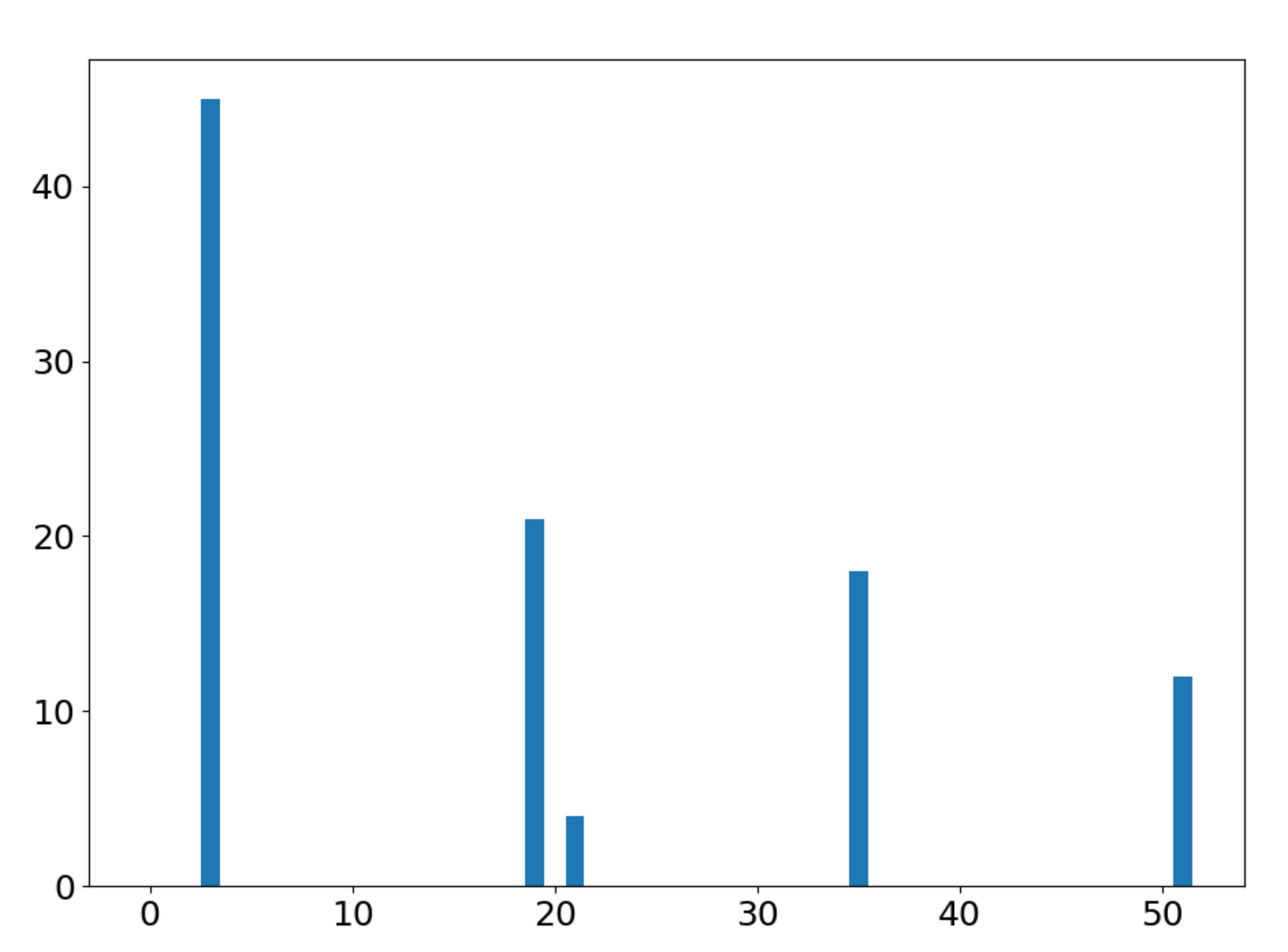}
    \setlength{\captionmargin}{10pt}
    \caption{\# of $n$-generation models w.r.t. $n_H$ for search (II) with $h^{1,1}=3$. The axes are the same as \figref{Higgshist1}.\\}
    \label{fig:Higgshist2}
\end{minipage}
\end{figure}

\begin{figure}[H]
    \centering
    \includegraphics[width=7.2cm]{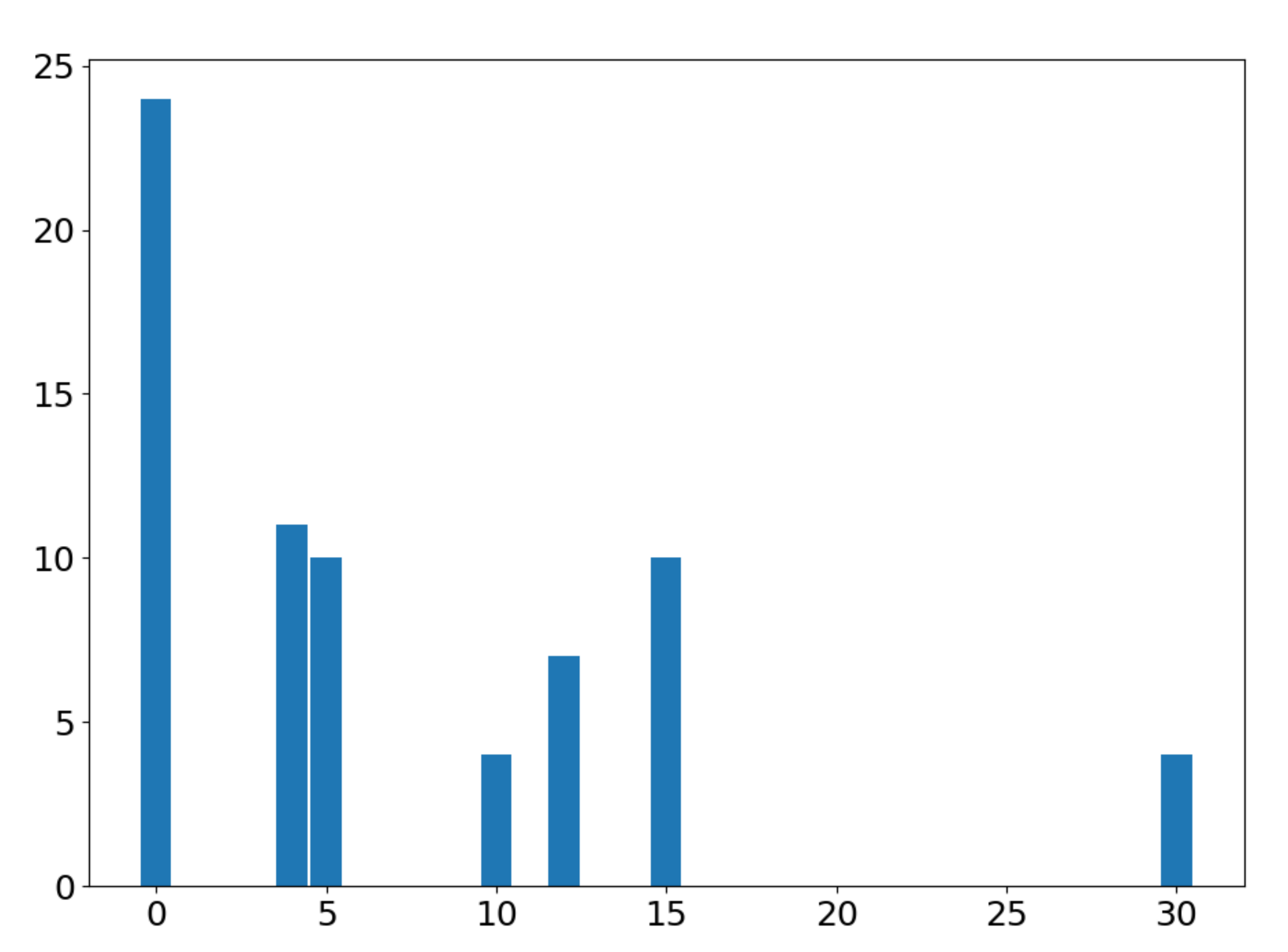}
    \caption{\# of $n$-generation models w.r.t. $n_H$ for search (III) with $h^{1,1}=3$. The axes are the same as \figref{Higgshist1}.}
    \label{fig:Higgshist3}
\end{figure}

In this section, we count the number of generations of Higgs (Higgsino) by implementing the analysis of 
the previous subsection. 
Note that we only take into account the Higgs pairs which are vector-like under the SM gauge group, but chiral with respect to other extra $U(1)$s. 
For definiteness, we restrict ourselves to $n_\phi\geq 0\:(\forall \phi\in\mathfrak{q}\cup\mathfrak{l}\cup\mathfrak{u}^c\cup\mathfrak{e}^c)$ cases and 
define the Higgs doublets from $\mathfrak{l}$ by checking the Yukawa couplings of quarks and leptons. 
In the obtained models, the generation number of up-type Higgs $n_{H_u}$ is same with 
that of down-type Higgs $n_{H_d}$. The condition $n_\phi\geq 0$ is so tight that only 3 cases (search (I), (II) and (III) with $h^{1,1}=3$) are able to be analyzed in our numerical analysis. 
Figs. \ref{fig:Higgshist1}, \ref{fig:Higgshist2} and \ref{fig:Higgshist3} show histograms of 
the number of Higgs pairs $n_H$ in the 3-generation island. 
From these figures, we find that $n_H$ listed in Table \ref{tab:Higgs generation} 
(except for $n_H=0$) is favored in the 3-generation island. 
Although there are not so many models in our limited search, it turns out that 1-pair Higgs models 
is disfavored. 
We expect that an existence of a large number of Higgs pairs is a generic property in heterotic string vacua. 
\begin{table}[H]
    \centering
    \begin{tabular}{|c|c|c|c|} \hline
        Search & $h^{1,1}$ & \# of $n_H>0$ models & Favored $n_H$\\ \hline
        (I) & 3 & 72 & 4,10,30 \\
        (II) & 3 & 100 & 3 \\
        (III) & 3 & 46 & 4 \\ \hline
    \end{tabular}
    \caption{Favored number of Higgs pairs $n_H$ except $n_H=0$ in the 3-generation island.}
    \label{tab:Higgs generation}
\end{table}

\section{Conclusions and Discussions}
\label{sec:conclusion}

In this paper, we applied the deep autoencoders 
and k-means++ clustering to the string landscape 
by employing the topological data of 
CY threefolds and internal gauge fluxes as input data. 
In particular, we investigated $SO(32)$ heterotic string vacua on smooth CICY threefolds with line bundles, taking into account the phenomenological and theoretical consistency conditions. 
After training the autoencoder on at most 161 input data, 
satisfying the consistency conditions as well as reproducing the $n$-generations of quarks and leptons without chiral exotics, 
we draw a 2D chart of the landscape of $n$-generation models by 
utilizing the k-means++ algorithm. 
It turned out that 3-generation models cluster in particular islands in the 2D chart 
and we called the cluster with densest three-generation models ``3-generation island". 
Such a structure has also been pointed out in the Mini-Landscape of heterotic $\mathbb{Z}_6$-II orbifold models~\cite{Mutter:2018sra}.\footnote{A decision tree employed in \cite{Mutter:2018sra} is applicable to our analysis which is valid to approximate the classification of models, but it is beyond our purpose.} 
We expect that the presence of 3-generation island will be a universal phenomena in the string landscape including $E_8\times E_8$ heterotic line bundle models as well as intersecting/magnetized D-brane models. 

By estimating the KL divergences of model parameters, we find that the clustered 3-generation island has a strong correlation with the topological 
data of CY threefolds, in particular, second Chern class of CY threefolds, namely $c_2(T{\cal M}) \in H^{2,2}({\cal M}, 18\mathbb{Z})$, although there is no bias in the second Chern class of CICYs we employed. 
It indicates that second Chern numbers of CYs provide a guideline to obtain 3-generation MSSM-like models. We leave to reveal the underlying reason for future work. 
It is interesting to apply our analysis to other regions of the string landscape and check the values of second Chern number of CYs for 3-generation models. 

We also counted the number of Higgs pairs which are vector-like under the SM gauge group, but chiral with respect to other extra $U(1)$s. Our results show that the 3-generation island contains a large number of Higgs pairs. 
It will motivate us to study the phenomenology of multi-Higgs models discussed in the bottom-up approach. 

Finally, we comment on possible applications of our method to other regions of the string landscape. 
It is straightforward to extend our analysis to $E_8\times E_8$ heterotic line bundle models by changing the gauge group decomposition.\footnote{See for the recent discussion of $E_8\times E_8$ heterotic line bundle models using deep reinforcement learning, Ref.~\cite{Larfors:2020ugo}.} 
For the D-brane models, it is enough to add the input data such as the position of D-branes, an amount of magnetic flux (in Type IIB magnetized D-brane models) 
and intersection angles (in Type IIA intersecting D-brane models), taking into account the proper tadpole cancellation conditions. We hope to report on this interesting work in the future.

\subsection*{Acknowledgements}

We would like to thank H. Abe, S. H. Lim and A. Otsuka for useful discussions. 
H. O. was supported in part by JSPS KAKENHI Grant Numbers JP19J00664 and JP20K14477.

\end{document}